\documentclass[%
reprint,
superscriptaddress,
amsmath,amssymb,
aps,
prx,
longbibliography
]{revtex4-2}
\usepackage{color}
\usepackage{graphicx}
\usepackage{dcolumn}
\usepackage{bm}
\usepackage{textcomp}

\usepackage{braket}

\usepackage{mathrsfs}
\usepackage{tikz}
\usetikzlibrary{arrows.meta,calc}

\begin {document}

\title{Breakdown of Linear Response Induced by Velocity-Dependent Stochastic Resetting}

\author{Yuto Takeishi}
\affiliation{%
  Department of Physics and Astronomy, Tokyo University of Science, Noda, Chiba 278-8510, Japan
}%

\author{Takuma Akimoto}
\email{takuma@rs.tus.ac.jp}
\affiliation{%
  Department of Physics and Astronomy, Tokyo University of Science, Noda, Chiba 278-8510, Japan
}%



\date{\today}
\begin{abstract}
Linear response theory lies at the foundation of transport phenomena, predicting that physical systems respond proportionally to weak external forces. Here we show that this principle can break down in a minimal nonequilibrium setting due to state-dependent stochastic resetting. We consider a driven Langevin particle subject to a resetting mechanism whose rate grows as a power of the particle velocity, motivated by transport processes where faster carriers experience more frequent scattering events.
We derive the exact steady-state velocity distribution and establish a moment balance relation that links external driving, viscous dissipation, and resetting-induced dissipation. This relation reveals that the response is controlled by a nonlinear coupling between the velocity and the resetting rate, leading to nonlinear transport. In particular, the mean velocity obeys the exact power law $\langle v\rangle \propto F^{1/(\alpha+1)}$, where $\alpha$ characterizes the velocity dependence of the resetting rate.
Our results provide a solvable example in which linear response fails at the level of the leading-order behavior and identify velocity-dependent resetting as a minimal dynamical mechanism for generating nonlinear transport in nonequilibrium steady states.
\end{abstract}

\maketitle


\section{Introduction}

Linear response theory provides a universal framework for understanding how physical systems react to weak external perturbations \cite{Onsager1931-I, *Onsager1931-II, kubo1966fluctuation,kubo2012statistical}. Classical examples include Ohm's law, Hooke's law, and Fourier's law, which express a proportionality between current and force, stress and strain, or heat flux and temperature gradient, respectively, and can be justified from microscopic dynamics under near-equilibrium conditions \cite{onsager1934deviations,Chernov1993,bonetto2000fourier}. These laws are deeply rooted in near-equilibrium statistical mechanics, where time-reversal symmetry and detailed balance ensure such proportionality \cite{Onsager1931-I, *Onsager1931-II, kubo1966fluctuation,kubo2012statistical}. 

Beyond equilibrium, however, the validity of linear response is more subtle. 
Remarkably, linear response often survives even in nonequilibrium steady states. 
Generalized fluctuation-response relations that separate entropic and frenetic contributions demonstrate that detailed balance is not a necessary condition for linear response to hold~\cite{baiesi2009nonequilibrium,Baiesi2009,Baiesi2013}. 
Likewise, deterministic nonequilibrium models such as thermostatted Lorentz gases exhibit linear response under weak driving despite operating far from equilibrium~\cite{dorfman1999introduction,Bonetto2002,j2007statistical}. 
Even in anomalous transport models such as continuous-time random walks, where diffusion becomes highly nontrivial, the response to weak external fields often remains linear~\cite{He2008,Akimoto2012einstein,Metzler2014}. 
These results suggest that linear response is surprisingly robust.

Nonlinear response beyond the linear-response regime has long been studied in nonequilibrium statistical mechanics~\cite{bochkov1981nonlinear,habdas2004forced,Lippiello2008,Gazuz2009,latorre2013corrections,Benichou2014,Benichou2016,sakai2025symmetry}. 
A paradigmatic example is provided by driven interacting systems such as driven lattice gases~\cite{katz1984nonequilibrium,derrida2002large,krapivsky2010kinetic}, 
where the combination of external driving and particle interactions produces nonequilibrium steady states with collective behavior, phase transitions, and nontrivial transport properties. 
In such systems, deviations from linear response are typically attributed to interactions and emergent correlations.
Nonlinear transport has also been reported in soft and disordered systems, including active matter and dense granular flows~\cite{berthier2002nonequilibrium, Fuchs2002,tailleur2009sedimentation,Fodor2016,pouliquen2009non,Kamrin2012},
as well as  quenched trap models~\cite{bouchaud90,Akimoto-Saito2020,burov2020transient,shafir2022case,Shafir2024}.
In these cases, nonlinear response is commonly associated with spatial heterogeneity, memory effects, or other forms of dynamical complexity.

Taken together, existing examples suggest that nonlinear response typically emerges from interactions, disorder, non-Markovian dynamics, or additional constraints. 
This raises a fundamental conceptual question:
\emph{What is the minimal dynamical ingredient required to destroy linear response, even under weak external forcing?}

A natural candidate for exploring this question is stochastic resetting, which provides a minimal and analytically tractable mechanism for generating nonequilibrium steady states. 
In resetting dynamics, a system is intermittently returned to a reference configuration, thereby explicitly breaking detailed balance~\cite{Evans2011,evans2011diffusion,Reuveni2016,Pal2017,fuchs2016stochastic,Chechkin2018,evans2020stochastic}. 
Despite this strong nonequilibrium character, most existing studies consider constant resetting rates and typically report linear response under weak driving~\cite{Sokolov2023,keidar2024universal}. 
It therefore remains unclear whether resetting alone---without interactions, disorder, or strong forcing---can fundamentally alter the structure of linear response.

In this work, we investigate a class of resetting dynamics in which the resetting (or scattering) rate depends explicitly on the instantaneous velocity of the particle. 
This generalizes the classical Drude picture of transport~\cite{Drude1900}, where scattering events occur at a constant rate independent of the carrier velocity. 
Within the Drude framework, this velocity-independent scattering is precisely what guarantees linear response at weak driving. 
Here we relax this assumption and allow the scattering rate to scale as a power law of the velocity. 
Such state-dependent scattering is physically natural: in transport models of electric conduction, faster carriers typically experience enhanced collision frequencies~\cite{bonetto2013nonequilibrium}, while in subrecoil laser cooling the photon scattering rate decreases with atomic momentum, leading to long-lived trapping near zero velocity~\cite{Bardou2002,Cohen-Tannoudji1998,Aspect1988,Barkai2021,barkai2022gas,Akimoto2022infinite}. 

We demonstrate that velocity-dependent resetting constitutes a minimal and exactly solvable mechanism for the breakdown of linear response.
Even arbitrarily weak external forcing leads to intrinsically nonlinear scaling, with
$\langle v \rangle \propto F^{1/(\alpha+1)}$,
showing that nonlinearity can emerge at leading order without interactions, disorder, or memory effects.

\section{State-Dependent Stochastic Reseting in Driven Systems}

We consider a particle moving along a one-dimensional axis, subject to a constant external force \( F \) and stochastic resetting of its velocity. The dynamics are governed by the Langevin equation:
\begin{equation}
  m\frac{dv}{dt} = F - \gamma v + \xi(t),
\end{equation}
where \( F \ge 0 \) is the external force, \( \gamma \ge 0 \) is the friction coefficient, and \( \xi(t) \) is a Gaussian white noise with zero mean and delta-correlated fluctuations:
\begin{equation}
  \langle \xi(t) \rangle = 0, \qquad \langle \xi(t)\xi(t') \rangle = 2D\,\delta(t - t'),
\end{equation}
where $D$ is a positive parameter controlling the noise intensity. If the system is coupled to a thermal bath, \( D \) may be interpreted as \( D = \gamma k_{\rm B} T \) via the fluctuation-dissipation relation, but we treat it as an independent parameter throughout this work.
For simplicity, we rescale units such that the particle mass is unity, \( m = 1 \). The Langevin equation then reduces to
\begin{equation}
  \frac{dv}{dt} = F - \gamma v + \xi(t).
\end{equation}
The dimensional form can be recovered by the substitutions \( F \rightarrow F/m \), \( \gamma \rightarrow \gamma/m \), and \( D \rightarrow D/m^2 \), if needed.

Extensions of stochastic resetting with time- or state-dependent rates have been explored in various stochastic processes~\cite{pal2016diffusion,evans2020stochastic}. 
Motivated by transport phenomena in which scattering events depend on the particle velocity~\cite{bonetto2013nonequilibrium}, we consider a resetting rate that depends explicitly on the instantaneous velocity. 
Specifically, when the particle has velocity \(v\), resetting occurs at a rate
\begin{equation}
  r(v) = \lambda |v|^{\alpha}, \qquad \lambda > 0,
\end{equation}
where \( \alpha \in \mathbb{R} \) is a tunable exponent characterizing the degree of the velocity dependence, and $\lambda$ sets the overall scale of the resetting rate. 
Power-law velocity dependences of scattering rates also arise in subrecoil laser cooling, where the photon scattering rate decreases with atomic momentum, corresponding to a positive exponent (\(\alpha>0\))~\cite{Bardou2002}. 
Although the scattering events in that system do not reset the velocity and instead modify the internal state of the atom, a similar functional dependence on velocity appears.

For \( \alpha > 0 \), fast-moving particles are more likely to reset, whereas for \( \alpha < 0 \), slow-moving ones reset more frequently. At each resetting event, the particle's velocity is instantaneously set to zero:  $v \rightarrow 0$.
The dynamics therefore alternates between two processes: continuous evolution under the Langevin dynamics---acceleration due to the external force together with friction and noise---and stochastic resetting of the velocity to zero at rate \( r(v) \). 
The competition between these processes drives the system toward a nonequilibrium stationary state with a well-defined velocity distribution.

To clarify the behavior of the stationary velocity distribution and the mean velocity as functions of the external force \( F \),  
we analyze the following representative cases in a hierarchical manner. 
We begin with a minimal deterministic model without friction or noise (\(\gamma = 0, D = 0\)), which can be solved analytically for a general resetting exponent \( \alpha \). This minimal setting allows us to isolate the fundamental role of state-dependent resetting.
We then turn to the general stochastic model including both friction and noise (\(\gamma > 0, D > 0\)), for which we focus on the physically motivated case \( \alpha = 1 \). Within this general framework, we further consider two limiting cases: dissipative dynamics with friction but no noise (\(\gamma > 0, D = 0\)), and diffusive dynamics without friction (\(\gamma = 0, D > 0\)). 
In addition, we examinethe special case \( \alpha = 0 \), which corresponds to constant-rate resetting and recovers linear response.
This organization allows us to first identify the fundamental effects of velocity-dependent resetting in a minimal setting, and then systematically examine how dissipation and diffusion modify the response in more realistic situations.

\section{Nonlinear Response in Representative Regimes}

Building on the classification above, we now analyze the stationary velocity distribution and the mean velocity for each representative case. Starting from the minimal deterministic model, we progressively introduce friction and noise to investigate how these ingredients shape the system's response to the external force.

\subsection{Minimal model and general nonlinear response}

We begin by analyzing the simplest case, in which both friction and noise are absent ($\gamma=0, D=0$). 
In this deterministic limit, stochastic resetting acts on a purely driven motion, a situation that has been considered previously in related contexts~\cite{eule2016non}. 
The equation of motion then reduces to
\begin{equation}
  \frac{dv}{dt} = F.
\end{equation}
Setting \( t = 0 \) as the initial time---either at the beginning of the trajectory or immediately after a reset---the particle's velocity increases linearly with time:
\begin{equation}
  v(t) = F t.
\end{equation}
The velocity is stochastically reset to zero at a rate that depends on the instantaneous velocity, i.e., $r(v) = \lambda v^{\alpha}$. 
Since the velocity is always non-negative (\( v(t) \ge 0 \)), the absolute value can be omitted. When a reset occurs, the velocity is instantaneously set to zero, and the particle resumes deterministic acceleration according to the same equation of motion. Repeating this process leads to the emergence of a stationary velocity distribution.

\subsubsection{Stationary velocity distribution}

The time evolution of the velocity probability density function (PDF) $P(v,t)$ is governed by a master equation that incorporates deterministic acceleration due to the external force, stochastic resetting events occurring at a velocity-dependent rate $r(v)$, and reinjection at the reset state $v=0$. 
Within the general framework of stochastic resetting, such dynamics can be described by a master equation containing a state-dependent loss term and a reinjection term at the reset configuration~\cite{evans2020stochastic}. 
Specifically,

\begin{widetext}
\begin{equation}
\frac{\partial P(v,t)}{\partial t}
= -\frac{\partial}{\partial v} \left( F P(v,t) \right) - r(v) P(v,t) 
+ \delta(v) \int_0^{\infty} r(v') P(v',t)\, dv'.
\end{equation}
\end{widetext}
In the long-time limit, the system reaches a stationary state where $\partial_t P(v,t) = 0$. Denoting the stationary distribution as $P_{\mathrm{st}}(v)$, the master equation reduces to
\begin{equation}
  \frac{d}{dv} \left( F P_{\mathrm{st}}(v) \right) = - r(v)\, P_{\mathrm{st}}(v) + \delta(v) \int_{0}^{\infty} r(v')\, P_{\mathrm{st}}(v')\, dv'.
\end{equation}
Here, the left-hand side captures the divergence of the probability flux due to acceleration by the external force, while the right-hand side consists of a loss term due to resetting and a source term at \( v = 0 \) from accumulated resets.

Since $P_{\mathrm{st}}(v)=0$ for $v<0$, the equation for \( v > 0 \) reduces to the ordinary differential equation
\begin{equation}
  F \frac{d P_{\mathrm{st}}(v)}{dv} = - r(v)\, P_{\mathrm{st}}(v) = - \lambda v^{\alpha} P_{\mathrm{st}}(v).
\end{equation}
Solving this equation yields
\begin{equation}
  P_{\mathrm{st}}(v) = C \exp\!\left[ - \frac{\lambda}{F(\alpha+1)} v^{\alpha+1} \right], \qquad (v>0)
\end{equation}
where $C$ is a normalization constant. The normalization condition
\begin{equation}
  \int_{0}^{\infty} P_{\mathrm{st}}(v)\, dv = 1
\end{equation}
gives
\begin{equation}
  C = \left( \frac{\lambda}{F(\alpha+1)} \right)^{\frac{1}{\alpha+1}} \frac{1}{\Gamma\!\left( \frac{\alpha+2}{\alpha+1} \right)} .
\end{equation}
Thus, the stationary velocity distribution for $v \ge 0$ takes the form
\begin{equation}
  P_{\mathrm{st}}(v) = \left( \frac{\lambda}{F(\alpha+1)} \right)^{\frac{1}{\alpha+1}} \dfrac{\exp\!\left[ - \frac{\lambda}{F(\alpha+1)} v^{\alpha+1} \right]}{\Gamma\!\left( \frac{\alpha+2}{\alpha+1} \right)} .
  \label{eq: st dist det}
\end{equation}
The normalization constant is finite only if \( \alpha > -1 \), so a normalizable stationary state exists exclusively in this regime. For \( \alpha \le -1 \), the resetting rate diverges as \( v \to 0 \) and no stationary distribution can be established. The resulting non-normalizable density is known as an \emph{infinite invariant density} in dynamical systems~\cite{Aaronson1997}, and arises in contexts such as anomalous diffusion and subrecoil laser cooling \cite{Akimoto2010, Akimoto2012, Leibovich2019, Akimoto2020-inf, Barkai2021, *barkai2022gas, Akimoto2022infinite}. 
In the remainder of this work, we restrict our analysis to the regime \( \alpha > -1 \), where a well-defined steady state exists.

\begin{figure*}
\includegraphics[width=.98\linewidth, angle=0]{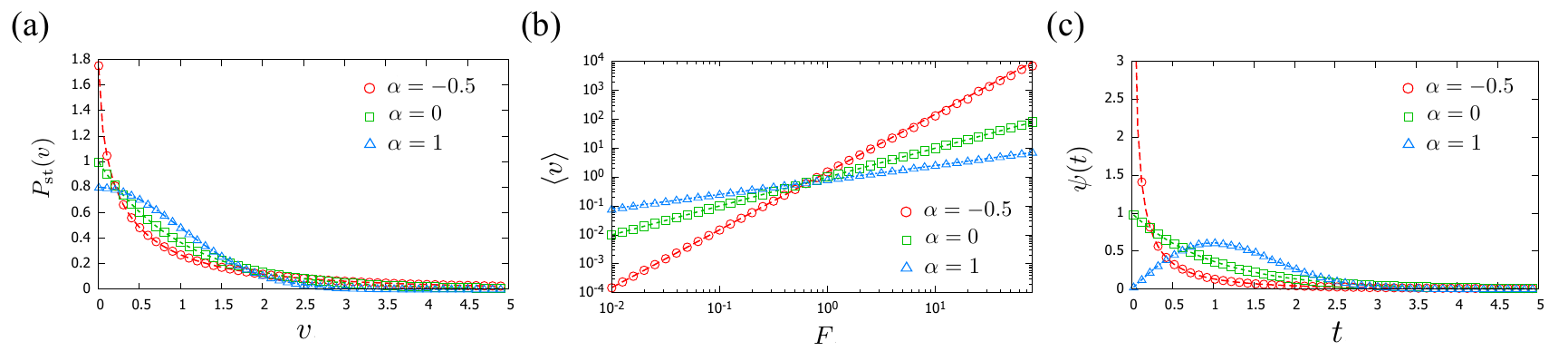}
\caption{Statistical properties of the minimal deterministic model with velocity-dependent resetting for different values of the resetting exponent $\alpha=-0.5, 0,$ and 1.
(a) Stationary velocity PDFs $P_{\mathrm{st}}(v)$ ($F=1$). Symbols denote numerical results, and dahsed lines show the exact analytical expression given by Eq.~\eqref{eq: st dist det}.
(b) Mean velocity \( \langle v \rangle \) as a function of the external force \( F \). Symbols denote numerical results, while dashed lines represent the theoretical prediction from Eq.~\eqref{eq: mean v case 1}. The slopes are consistent with the exact scaling exponent \( 1/(\alpha+1) \), demonstrating nonlinear response for \( \alpha \neq 0 \) and recovery of linear response for \( \alpha = 0 \).
(c) Inter-reset-time distribution $\psi(t)$  ($F=1$). Symbols denote numerical results and dashed lines correspond to the analytical expression in Eq.~\eqref{eq: waiting-time dist det}. The dependence on $\alpha$ illustrates how velocity-dependent resetting modifies the reset statistics.}
\label{fig: stat. prop. deterministic}
\end{figure*}

Figure~\ref{fig: stat. prop. deterministic}(a) compares the stationary velocity PDFs $P_{\mathrm{st}}(v)$ for $\alpha=-0.5,0, 1$. The numerical results (symbols) agree excellently with the exact prediction in Eq.~\eqref{eq: st dist det} (solid lines). For all $\alpha$, the PDF is peaked at the resetting state $v=0$, reflecting the continuous injection of probability at the origin.
In the special case $\alpha=0$, corresponding to constant-rate (Poissonian) resetting, the stationary PDF reduces to a purely exponential form, which serves as a useful reference profile.
For $\alpha<0$, the resetting rate is larger at small velocities, leading to an enhanced accumulation of probability near the origin and a sharper peak. In contrast, for $\alpha>0$, resetting preferentially suppresses high-velocity states, producing a faster decay of the distribution away from $v=0$.

\subsubsection{Mean velocity}

The mean velocity in the steady state is given by
\begin{equation}
  \langle v \rangle = \left( \frac{\alpha+1}{\lambda} \right)^{\!\frac{1}{\alpha+1}} \frac{\Gamma\!\left( \frac{2}{\alpha+1} \right)} {\Gamma\!\left( \frac{1}{\alpha+1} \right)} F^{\frac{1}{\alpha+1}}.
  \label{eq: mean v case 1}
\end{equation}
which reveals an exact power-law response, $\langle v \rangle \propto F^{\beta}$ with $\beta = \frac{1}{\alpha+1}$.
Importantly, this nonlinear scaling holds for all values of $F$, not merely as an asymptotic limit.

Figure~\ref{fig: stat. prop. deterministic}(b) confirms this prediction. The numerical results (symbols), obtained from direct simulations of the deterministic dynamics with stochastic resetting, are in excellent agreement with the exact analytical expression (solid lines). In the log-log representation, the slopes clearly match the exponent $1/(\alpha+1)$ for each value of $\alpha$, demonstrating the validity of the exact scaling law.

The sign of $\alpha$ determines whether the response is suppressed or enhanced relative to the linear case.
For $\alpha > 0$, the resetting rate increases with velocity, leading to a sublinear response ($\beta < 1$).
For $-1 < \alpha < 0$, resetting is stronger at small velocities, resulting in a superlinear response ($\beta > 1$).
The special case $\alpha = 0$ corresponds to constant-rate resetting and recovers linear response, $\langle v \rangle = F/\lambda$.
Notably, for $\alpha \neq 0$, the deviation from linear response is not a higher-order correction at large driving but persists down to arbitrarily small forces, indicating a genuine breakdown of linear response induced solely by velocity-dependent resetting.

\subsubsection{Inter-reset-time distribution}

To deepen our understanding of the system's response to the external force, we examine the inter-reset-time statistics between successive reset events, which quantify how frequently resetting occurs in the steady state. 
In contrast to self-exciting processes such as Hawkes processes, where the hazard rate itself is stochastic and history-dependent~\cite{ogata1981lewis,daw2018queues}, the present model retains a renewal structure: after each reset, the dynamics restarts from the same initial velocity. The hazard rate is therefore fully determined by the deterministic trajectory $v(t)$ between resets.
Since the reset rate depends on the instantaneous velocity $v(t)$, it is inherently time-dependent through the trajectory of the particle.
The PDF  \( \psi(t) \) for the first reset is given by
\begin{equation}
  \psi(t) = \lambda v^{\alpha}(t) \exp\left[ - \int_{0}^{t} \lambda v^{\alpha}(t')\, dt' \right].
\end{equation}
This expression is obtained as the product of the instantaneous reset rate at time $t$, 
$\lambda |v(t)|^{\alpha}$, and the survival probability, i.e., the probability that no reset has occurred up to time $t$, given by the exponential factor. 
Although the overall dynamics is Markovian, the inter-reset times are generally not exponentially distributed. 
Consequently, the resetting events do not constitute a homogeneous Poisson process. 
Rather, the process can be viewed as a Markov dynamics with a state-dependent killing rate~\cite{DaleyVereJones2003}, 
where the effective resetting intensity varies along the deterministic trajectory of the particle. 
In the present model, $v(t)$ evolves deterministically between resets under the applied force (e.g., $v(t)=Ft$ in the frictionless case).
Substituting this into the expression yields
\begin{equation}
  \psi(t) = \lambda (F t)^{\alpha} \exp\left[ - \frac{\lambda F^{\alpha}}{\alpha + 1} t^{\alpha + 1} \right].
  \label{eq: waiting-time dist det}
\end{equation}
The mean waiting time \( \langle \tau \rangle \) is then obtained as
\begin{equation}
  \langle \tau \rangle = \frac{1}{F} \left( \frac{\alpha + 1}{\lambda} \right)^{\frac{1}{\alpha + 1}} \Gamma\left( \frac{\alpha + 2}{\alpha + 1} \right).
\end{equation}
This quantity depends explicitly on the resetting strength $\lambda$ and the exponent $\alpha$. Since $\langle \tau \rangle$ characterizes the typical duration of acceleration before a reset occurs, it directly determines how far the particle can move under the external force.
When $\alpha > 0$, fast particles reset more frequently, leading to shorter waiting times and thus more frequent interruptions in acceleration. This results in a suppressed mean velocity. Conversely, when $\alpha < 0$, slow particles reset more often, allowing fast particles to continue accelerating longer. The resulting increase in $\langle \tau \rangle$ enhances transport.
Therefore, the impact of $\alpha$ on the response behavior can be quantitatively understood in terms of the mean waiting time it induces.

Figure~\ref{fig: stat. prop. deterministic}(c) shows the numerical results for the mean velocity as a function of the external force \( F \), for three representative values of the exponent: \( \alpha = -0.5 \), \( 0 \), and \( 2 \). According to Eq.~\eqref{eq: mean v case 1}, the theoretical scaling predicts \( \langle v \rangle \propto F^2 \) for \( \alpha = -0.5 \), linear response for \( \alpha = 0 \), and \( \langle v \rangle \propto F^{1/3} \) for \( \alpha = 2 \), all of which are confirmed by the numerical data.  
These results quantitatively validate the theoretical prediction that smaller values of \( \alpha \) lead to stronger response, while larger \( \alpha \) suppress transport due to more frequent resetting of fast-moving particles.

In the followings, we focus on the case $\alpha = 1$, for which the resetting rate is linearly proportional to the velocity. This choice is not only analytically tractable but also physically motivated: velocity-dependent resetting with $\alpha = 1$ naturally arises in systems where faster particles are more likely to encounter scattering or dissipation.

\subsection{General dynamics with friction and noise ($\gamma>0$, $D>0$) for $\alpha=1$}

In this section, we analyze a general model that includes both friction and noise ($\gamma>0$, $D>0$).
The particle’s velocity evolves according to the Langevin equation
\begin{equation}
\frac{dv}{dt} = F - \gamma v + \xi(t),
\end{equation}
where $\xi(t)$ is Gaussian white noise with $\langle \xi(t)\xi(t') \rangle = 2D\,\delta(t-t')$.
This corresponds to the general case with both friction and noise ($\gamma > 0, D > 0$).
Due to the presence of friction, the particle velocity experiences acceleration induced by the external force and deceleration due to friction, while simultaneously undergoing diffusion in velocity space driven by thermal noise.
Stochastic resetting occurs with a velocity-dependent rate given by
\begin{equation}
r(v) = \lambda |v| \qquad (\lambda>0).
\end{equation}
This form implies that resetting events are more likely to occur at larger velocities.
Whenever a resetting event takes place, the particle velocity is instantaneously set to $v=0$.
Subsequently, the particle resumes its stochastic acceleration under the influence of the external force and noise.
Through the repetition of this process, a nonequilibrium steady-state velocity distribution is established.

\subsubsection{Stationary velocity distribution}

The time evolution of the velocity PDF $P(v,t)$ is governed by a master equation in which resetting acts as a velocity-dependent 
loss term with reinjection at the reset state~\cite{evans2020stochastic}.

\begin{widetext}
\begin{equation}
\frac{\partial P(v,t)}{\partial t}
=  -\frac{\partial}{\partial v}\!\left[(F-\gamma v) P(v,t)\right] + D \frac{\partial^2 P(v,t)}{\partial v^2} - r(v) P(v,t) + \delta(v) \int_{0}^{\infty} r(v') P(v',t)\, dv'.
\label{eq: master eq. general}
\end{equation}
In the stationary state $\partial_t P = 0$, the velocity distribution \( P_{\mathrm{st}}(v) \) satisfies
\begin{equation}
\frac{d}{dv}\!\left[(F-\gamma v) P_{\rm st}(v)\right] = D \frac{d^2 P_{\rm st}(v)}{dv^2} - r(v) P_{\rm st}(v) + \delta(v) \int_{-\infty}^{\infty} r(v') P_{\rm st}(v')\,dv'.
\label{eq: stationary equation general}
\end{equation}
The left-hand side represents the change in probability current due to the external force $F$ and friction $\gamma$.
On the right-hand side, the first term describes diffusion in velocity space, the second term accounts for the loss of probability due to resetting at velocity $v$, and the third term is a source term corresponding to probability injection at $v=0$ induced by resetting.

For $v\neq 0$, the delta-function term vanishes, and the equation reduces to
\begin{equation}
\frac{d}{dv}\!\left[(F-\gamma v) P_{\rm st}(v)\right] = D \frac{d^2 P_{\rm st}(v)}{dv^2} - \lambda |v| P_{\rm st}(v).
\end{equation}
Since the resetting rate is given by $r(v)=\lambda |v|$, the equation must be treated separately for $v>0$ and $v<0$.
For $v>0$, the steady-state distribution satisfies
\begin{equation}
D P_{\rm st}''(v) - (F-\gamma v) P_{\rm st}'(v) + (\gamma-\lambda v) P_{\rm st}(v) = 0,
\end{equation}
where the prime denotes differentiation with respect to $v$.
To eliminate the first-derivative term, we perform the transformation
\begin{equation}
P_{\rm st}(v) = \exp\!\left(\frac{2Fv-\gamma v^2}{4D}\right) u(v).
\end{equation}
Substituting this expression into the differential equation yields
\begin{equation}
u''(v) + \left[ \frac{\gamma}{2D} - \frac{(F-\gamma v)^2}{4D^2} - \frac{\lambda v}{D} \right] u(v) = 0.
\end{equation}
Completing the square in the quadratic form of $v$, this equation can be rewritten as
\begin{equation}
u''(v) - \frac{\gamma^2}{4D^2} \left( v - \frac{F}{\gamma} + \frac{2\lambda D}{\gamma^2} \right)^2 u(v) + \left( \frac{\gamma}{2D} - \frac{\lambda F}{\gamma D} + \frac{\lambda^2}{\gamma^2} \right) u(v) = 0.
\end{equation}
Introducing the rescaled variable
\begin{equation}
z = \sqrt{\frac{\gamma}{D}} \left( v - \frac{F}{\gamma} + \frac{2\lambda D}{\gamma^2} \right),
\end{equation}
the above equation is reduced to the standard parabolic cylinder equation
\begin{equation}
\frac{d^2 u}{dz^2} + \left( \nu_+ + \frac{1}{2} - \frac{z^2}{4} \right) u = 0,
\end{equation}
with
\begin{equation}
\nu_+ = \frac{\lambda^2 D}{\gamma^3} - \frac{\lambda F}{\gamma^2}.
\end{equation}
The general solution is given by
\begin{equation}
u(z) = A \mathscr{D}_{\nu_+}(z) + B \mathscr{D}_{\nu_+}(-z),
\end{equation}
where $\mathscr{D}_\nu(z)$ denotes the parabolic cylinder function.
Imposing the physical boundary condition $P_{\rm st}(v)\to 0$ as $v\to\infty$, the term proportional to $\mathscr{D}_{\nu_+}(-z)$ diverges, and hence we set $B=0$.
As a result, for $v>0$ the steady-state velocity distribution is given by
\begin{equation}
P_{\rm st}(v) = C_+ \exp\!\left(\frac{Fv}{2D}-\frac{\gamma v^2}{4D}\right) \mathscr{D}_{\nu_+}\!\left(\sqrt{\frac{\gamma}{D}}(v-v_+^*)\right), \qquad (v>0).
\label{eq: st dist deterministic v>0}
\end{equation}
A similar analysis for $v<0$ yields
\begin{equation}
P_{\rm st}(v) = C_- \exp\!\left(\frac{Fv}{2D}-\frac{\gamma v^2}{4D}\right) \mathscr{D}_{\nu_-}\!\left(-\sqrt{\frac{\gamma}{D}}(v-v_-^*)\right), \qquad (v<0),
\label{eq: st dist deterministic v<0}
\end{equation}
\end{widetext}
where
\begin{equation}
v_\pm^* = \frac{F}{\gamma} \mp \frac{2\lambda D}{\gamma^2}, \qquad \nu_- = \frac{\lambda^2 D}{\gamma^3} + \frac{\lambda F}{\gamma^2}.
\end{equation}
Here $C_+$ and $C_-$ are normalization constants in the respective velocity domains.
The constants $C_+$ and $C_-$ are determined from the continuity condition at $v=0$,
\begin{equation}
P_{\rm st}(0^+) = P_{\rm st}(0^-),
\end{equation}
together with the normalization condition
\begin{equation}
\int_{-\infty}^{\infty} P_{\rm st}(v)\,dv = 1.
\end{equation}
These conditions yield
\begin{equation}
C_+ = \frac{A_-}{A_- J_+ + A_+ J_-}, \qquad C_- = \frac{A_+}{A_- J_+ + A_+ J_-},
\end{equation}
where
\begin{equation}
A_\pm = \mathscr{D}_{\nu_\pm}\!\left(\sqrt{\frac{\gamma}{D}}(-v_\pm^*)\right),
\end{equation}
and
\begin{equation}
J_\pm = \int_0^{\infty} \exp\!\left(\frac{Fv}{2D}-\frac{\gamma v^2}{4D}\right) \mathscr{D}_{\nu_\pm}\!\left(\pm\sqrt{\frac{\gamma}{D}}(v-v_\pm^*)\right)\,dv.
\end{equation}
The stationary velocity distribution for the general dynamics
is determined by the balance between drift, diffusion,
and velocity-dependent resetting.

\begin{figure*}
\includegraphics[width=.98\linewidth, angle=0]{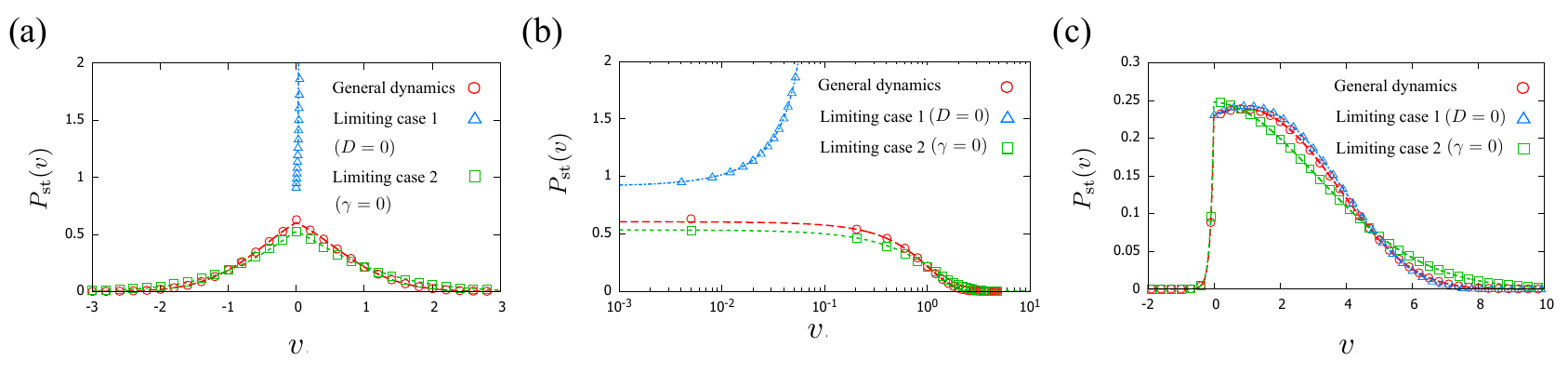}
\caption{Stationary velocity distributions $P_{\mathrm{st}}(v)$ for the general dynamics with $\alpha = 1$. 
Symbols denote numerical results, while dashed lines represent the analytical expressions given by 
Eqs.~\eqref{eq: st dist deterministic v>0} and \eqref{eq: st dist deterministic v<0}. 
The general dynamics ($D=1$, $\gamma=1$) is shown together with the limiting cases 
$D=0$ ($\gamma=1$) and $\gamma=0$ ($D=1$). 
(a) Distributions under a weak external force $F=0.1$. 
(b) Same data as in (a), shown for $v>0$ with a logarithmic scale on the vertical axis. 
(c) Distributions under a strong external force $F=10$.
}
\label{fig: stat. dist. general}
\end{figure*}

To illustrate the resulting behavior, Fig.~\ref{fig: stat. dist. general} shows the stationary velocity distributions obtained from direct numerical simulations for several parameter sets, including the limiting cases $D=0$ and $\gamma=0$, indicating that the qualitative structure of the stationary distribution
is largely controlled by the resetting mechanism.
The analytical expressions for these limiting cases will be presented in the following sections.

In the weak-driving regime [Fig.~\ref{fig: stat. dist. general}(a)], the distribution remains nearly symmetric around $v=0$, reflecting the weak bias introduced by the external force. The numerical results for the general dynamics ($D=1,\gamma=1$) lie between the two limiting cases $D=0$ and $\gamma=0$, indicating that both diffusion and friction broaden the velocity distribution while preserving its overall structure.

In the strong-driving regime [Fig.~\ref{fig: stat. dist. general}(b)], the stationary distribution becomes strongly skewed toward positive velocities. This asymmetry reflects the competition between the constant acceleration induced by the external force and the resetting mechanism that intermittently returns the velocity to zero. Despite the presence of noise and friction in the general dynamics, the numerical results remain consistent with the analytical structure predicted for the limiting cases, indicating that the qualitative shape of the stationary distribution is primarily governed by the resetting dynamics.

\paragraph{Limiting Case 1: Dissipative dynamics with friction but no noise.}

We now analyze a special case where thermal noise is absent, and the dynamics are governed solely by friction and the external force. The velocity of the particle evolves according to
\begin{equation}
  \frac{dv}{dt} = F - \gamma v,
\end{equation}
where \( F \) denotes the strength of the external field and \( \gamma \) is the friction coefficient. Setting \( t = 0 \) at the initial time or immediately after a reset, the particle accelerates under the external force while being slowed down by friction.

The stationary distribution can be obtained by setting $D=0$ in Eq.~(\ref{eq: stationary equation general}). 
 For \(v<0\), the stationary distribution vanishes, \(P_{\mathrm{st}}(v)=0\). For \(v>0\), the master equation reduces to
\begin{equation}
  \frac{d}{dv}\!\left[(F-\gamma v)\, P_{\mathrm{st}}(v)\right] = -\lambda v\, P_{\mathrm{st}}(v).
\end{equation}
where we have substituted \( r(v) = \lambda v \). 
Solving this equation yields
\begin{equation}
  P_{\mathrm{st}}(v) = C\, (F-\gamma v)^{\lambda F/\gamma^{2}-1} \exp\!\left(\frac{\lambda}{\gamma}v\right)
\end{equation}
for $0<v<v_{\max}=F/\gamma$, where \(C\) is a normalization constant and \( v_{\max} \) is the upper bound imposed by friction, corresponding to the terminal velocity.
The normalization condition
\begin{equation}
  \int_{0}^{v_{\max}} P_{\mathrm{st}}(v)\, dv = 1
\end{equation}
determines the constant \( C \) as
\begin{equation}
  C = \frac{\gamma}{\Gamma(\theta F,\, \theta F)} (\theta F)^{\theta F} \exp\!\left(-\theta F\right),
\end{equation}
where \(\Gamma(a,b)\) is the lower incomplete Gamma function and we define 
$\theta := \lambda /\gamma^2$, which sets the overall scale of the velocity-dependent resetting rate relative to the friction. The product \( \theta F = \lambda F / \gamma^2 \) acts as the key parameter that controls the shape of the stationary distribution and determines the crossover between linear and nonlinear response.
The stationary velocity distribution in this case exhibits several notable features arising from the interplay between friction and velocity-dependent resetting. First, due to the presence of friction, the velocity is strictly bounded above by \( v_{\max} \), beyond which the deterministic drift cannot carry the particle. The distribution has a finite support over \( 0 < v < v_{\max}  \), unlike the unbounded case in the absence of friction.

The functional form of $P_{\mathrm{st}}(v)$ combines a power-law prefactor and an exponential term. 
For small $v$, the distribution is suppressed because the reset rate $r(v)=\lambda v$ vanishes at $v=0$, leading to a vanishing probability flux into small velocities. 
In contrast, near the upper edge $v=v_{\max}$, the prefactor $(F-\gamma v)^{\theta F-1}$ introduces either a divergence or suppression depending on the value of the dimensionless parameter $\theta F$. 
In particular, in the weak-field regime ($\theta F<1$) the stationary distribution diverges as $v \to v_{\max}$, whereas for $\theta F>1$ it vanishes there. 
This behavior is clearly visible in Fig.~\ref{fig: stat. dist. general}(b), where the distribution exhibits a pronounced peak near $v_{\max}$ in the weak-driving case.

This crossover reflects the competition between deterministic acceleration, which drives the velocity toward $v_{\max}$, and resetting events that repeatedly return the particle to $v=0$. 
The overall shape of the stationary distribution is therefore controlled by the effective dimensionless parameter $\theta F$, which sets the relative strength of resetting and deterministic drift. 
As this parameter increases, the distribution shifts toward lower velocities, indicating a stronger influence of resetting. 
In the limit $\lambda \to 0$, the distribution collapses to a delta peak at $v=v_{\max}$, as expected for deterministic motion without resetting.

\paragraph{Limiting Case 2: Diffusive dynamics with noise but no friction}

In contrast to the previous one, we analyze a model without friction ($\gamma = 0$), in which Brownian motion in velocity space is introduced instead ($D > 0$). 
The particle velocity evolves according to the stochastic equation of motion
\begin{equation}
  \frac{dv}{dt} = F + \xi(t),
\end{equation}
where \( \xi(t) \) is Gaussian white noise with intensity \( D \), as defined previously.
In the absence of friction, the velocity is monotonically accelerated by the external force, while simultaneously spreading due to Brownian fluctuations. Unlike the frictional case discussed in the previous section, the velocity does not approach a terminal value; instead, in the absence of resetting, it would increase without bound on average.


The stationary distribution can be obtained by setting $D=0$ in Eq.~(\ref{eq: stationary equation general}). 
For $v \neq 0$, the delta-function term vanishes, and the stationary distribution satisfies the ordinary differential equation
\begin{equation}
  \frac{d}{dv}\!\left[F P_{\mathrm{st}}(v)\right] = D \frac{d^2 P_{\mathrm{st}}(v)}{dv^2} - r(v) P_{\mathrm{st}}(v).
\end{equation}
Since the resetting rate is chosen as $r(v) = \lambda |v|$, the equation must be treated separately for $v>0$ and $v<0$. For $v>0$, it reduces to
\begin{equation}
  D P''(v) - F P'(v) - \lambda v P(v) = 0,
\end{equation}
where primes denote derivatives with respect to $v$. To eliminate the first-derivative term, we apply the transformation 
\begin{equation}
  P(v) = \exp\!\left(\frac{Fv}{2D}\right) u(v),
\end{equation}
which yields the following equation for $u(v)$:
\begin{equation}
  u''(v) - \frac{\lambda}{D} \left( v + \frac{F^2}{4D\lambda} \right) u(v) = 0.
\end{equation}
Introducing the rescaled variable
\begin{equation}
  z = \left(\frac{\lambda}{D}\right)^{1/3} \left( v + \frac{F^2}{4D\lambda} \right),
\end{equation}
the equation reduces to the standard Airy equation, 
\begin{equation}
  \frac{d^2 u}{dz^2} - z u = 0.
\end{equation}
The general solution is thus a linear combination of Airy functions:
\begin{equation}
  u(z) = A\,\mathrm{Ai}(z) + B\,\mathrm{Bi}(z).
\end{equation}
Here, $\mathrm{Ai}(z)$ and $\mathrm{Bi}(z)$ are the standard Airy functions, which form a fundamental set of solutions to the differential equation $u''(z) - z u = 0$.
Among them, $\mathrm{Ai}(z)$ decays to zero as $z \to +\infty$, while $\mathrm{Bi}(z)$ diverges in this limit.
Imposing the physical boundary condition $P_{\mathrm{st}}(v) \to 0$ as $v \to \infty$ excludes the $\mathrm{Bi}$ solution, which diverges in this limit, and thus $B=0$. Consequently, for $v>0$ the stationary distribution reads
\begin{equation}
  P_{\mathrm{st}}(v) = C_{+}\, \exp\!\left(\frac{Fv}{2D}\right) \mathrm{Ai}\!\left[ \left(\frac{\lambda}{D}\right)^{1/3} \left( v + \frac{F^2}{4D\lambda} \right) \right].
\end{equation}
A similar analysis for $v<0$ leads to
\begin{equation}
  P_{\mathrm{st}}(v) = C_{-}\, \exp\!\left(\frac{Fv}{2D}\right) \mathrm{Ai}\!\left[ \left(\frac{\lambda}{D}\right)^{1/3} \left( -v + \frac{F^2}{4D\lambda} \right) \right].
\end{equation}
The constants $C_{+}$ and $C_{-}$ are normalization factors for the respective regions. Imposing continuity of the stationary distribution at $v=0$,
\begin{equation}
  P_{\mathrm{st}}(0^{+}) = P_{\mathrm{st}}(0^{-}),
\end{equation}
one finds $C_{+} = C_{-}$. Therefore, the stationary distribution can be expressed in a unified form over the entire real axis as
\begin{equation}
  P_{\mathrm{st}}(v) = C\, \exp\!\left(\frac{Fv}{2D}\right) \mathrm{Ai}\!\left[ \left(\frac{\lambda}{D}\right)^{1/3} \left( |v| + \frac{F^2}{4D\lambda} \right) \right].
\end{equation}
The normalization constant $C$ is determined by the condition
\begin{equation}
  \int_{-\infty}^{\infty} P_{\mathrm{st}}(v)\, dv = 1.
\end{equation}

The stationary velocity distributions for the diffusive dynamics without friction (\(\gamma=0\)) are also shown in Fig.~\ref{fig: stat. dist. general}. 
Compared with the general dynamics discussed above, the absence of friction leads to a broader velocity distribution, since the deterministic relaxation toward a finite velocity scale is no longer present. 
Diffusion therefore plays the dominant role in spreading the velocities around the deterministic drift trajectory.
In the weak-driving regime [Fig.~\ref{fig: stat. dist. general}(a)], the distribution remains centered near \(v=0\), but is noticeably wider than in the general case due to the combined effects of diffusion and resetting. 
For strong driving [Fig.~\ref{fig: stat. dist. general}(b)], the distribution becomes strongly skewed toward positive velocities, similar to the general dynamics. 
However, the lack of friction allows the velocity to explore larger values before a reset occurs, resulting in a broader positive tail.

\subsubsection{Mean velocity}

To gain analytical insight into the nonlinear response behavior, we focus on the asymptotic regimes $F \to 0$ and $F \to \infty$. 
We begin by deriving an exact relation for the first velocity moment.
Multiplying the steady-state master equation, Eq.~(\ref{eq: stationary equation general}), by $v$ and integrate over the entire velocity space, 
 the boundary terms vanish and we obtain
\begin{equation}
\gamma \langle v\rangle = F - \lambda \langle v|v|\rangle .
\label{eq: relation mean v}
\end{equation}
Thus, in the steady state, the external driving is exactly balanced by viscous dissipation and velocity-dependent resetting.

We assume that, in the weak-field limit, the steady-state distribution and its moments can be expanded analytically in powers of the external field $F$.
Accordingly, we write
\begin{equation}
P_{\rm st}(v) = P_0(v) + F P_1(v) + O(F^2),
\end{equation}
and hereafter restrict ourselves to linear response, neglecting terms of order $O(F^2)$ and higher.
By symmetry, the zeroth-order distribution $P_0(v)$ is even in $v$, implying that the mean velocity vanishes at $F=0$. Consequently, any nonzero mean velocity must arise at linear order in $F$.

To obtain a tractable expression for the mixed moment $\langle v|v| \rangle$, we further introduce a shift approximation: in the weak-field regime, the steady-state distribution is assumed to retain its zero-field shape and be shifted by a small amount $\delta = \langle v \rangle$. That is,
\begin{equation}
P_{\rm st}(v) \simeq P_0(v-\delta) \simeq P_0(v) - \delta \frac{dP_0(v)}{dv}.
\end{equation}
This approximation is expected to hold when weak driving produces
only a small asymmetry without significantly distorting the
overall shape of the distribution.
However, in the absence of friction ($\gamma=0$), the external
driving can modify the functional form of the stationary
distribution itself, and the simple shift approximation may
therefore become inaccurate.
Using this approximation, we expand
\begin{equation}
(v+\delta)|v+\delta| = v|v| + 2\delta |v| + O(\delta^2),
\end{equation}
which yields
\begin{equation}
\langle v|v|\rangle \simeq 2 \langle |v| \rangle_0 \langle v\rangle,
\end{equation}
where $\langle |v| \rangle_0$ denotes the mean absolute velocity in the absence of the external field ($F=0$).
Substituting this relation into the moment balance equation leads to the weak-field estimate
\begin{equation}
\langle v\rangle \simeq \frac{F}{\gamma + 2 \lambda \langle |v| \rangle_0}.
\label{eq: mean velocity general alpha=1}
\end{equation}
Since the zero-field distribution $P_0(v)$ is expressed in terms of parabolic cylinder functions, the quantity $\langle |v| \rangle_0$ is generally not expressible in a simple closed analytic form, but it can be evaluated numerically.
This expression shows that, in the weak-field regime, the system obeys linear response with an effective friction coefficient
\begin{equation}
\gamma_{\rm eff}
=
\gamma + 2\lambda \langle |v| \rangle_0.
\end{equation}
Thus, velocity-dependent resetting contributes an additional dissipative term proportional to the typical magnitude of velocity fluctuations in the zero-field state. In this sense, resetting acts as an effective nonlinear friction mechanism that renormalizes the linear mobility of the particle. Importantly, the diffusion constant $D$ enters the response only through the zero-field moment $\langle |v| \rangle_0$, indicating that thermal fluctuations regularize the weak-field behavior without altering the linear structure of the response.

In the strong-field limit, the dynamics is dominated by the external force and the resetting mechanism, while the contributions from friction and diffusion become relatively small.
In this regime, the model asymptotically reduces to the deterministic case, yielding
\begin{equation}
\langle v\rangle \sim \sqrt{\frac{2 F}{\pi\lambda}}\, , \qquad (F\to\infty).
\end{equation}
 Notably, the scaling exponent 1/2 is independent of both the friction coefficient $\gamma$ and the diffusion constant $D$.
This indicates that the asymptotic transport behavior is governed solely by the competition between deterministic acceleration and velocity-dependent resetting.

Together with the linear-response result in the weak-field regime, this establishes a crossover from linear transport at small $F$ to universal square-root scaling at large $F$. 
The crossover force can be estimated by matching the weak- and strong-field asymptotics, yielding
\begin{equation}
F_c =
\frac{2 \gamma_{\rm eff}^2}{\pi\lambda}.
\end{equation}
This shows that while the asymptotic square-root exponent is universal, the location of the crossover depends on the effective friction set by dissipation and resetting.

\paragraph{Limiting Case 1: Dissipative dynamics with friction but no noise.}

In the Limiting Case 1, the stationary mean velocity is given by
\begin{equation}
  \langle v \rangle = \frac{\gamma (\theta F)^{\theta F}}{\lambda\, \Gamma(\theta F,\, \theta F)} e^{-\theta F }.
\end{equation}
In the weak-field limit \(F\to 0\), using the asymptotic expansion of the incomplete Gamma function, we obtain
\begin{equation}
  \langle v\rangle \sim \frac{F}{\gamma} .
  \label{eq: mv small F case 2}
\end{equation}
This indicates that the system obeys linear response in the weak-driving regime.
In the large $F$ limit, application of Laplace's method leads to
\begin{equation}
  \langle v\rangle \sim \sqrt{\frac{2F}{\pi\lambda}}.
  \label{eq: mv large F case 2}
\end{equation}
This establishes a crossover from linear response at small driving to a universal square-root scaling at large fields, induced purely by the velocity-dependent resetting mechanism.

The crossover between linear and nonlinear response occurs around a characteristic force \( F_c \), which is given by
\begin{equation}
  F_c = \frac{2 \gamma^2}{\pi \lambda}.
\end{equation}
For \( F \ll F_c \), the system exhibits linear response, while for \( F \gg F_c \), the mean velocity grows sublinearly as \( \langle v \rangle \propto \sqrt{F} \). 
Interestingly, a similar type of crossover behavior has also been observed in the quenched trap model, where nonlinear response emerges due to trapping-induced memory effects \cite{bouchaud90}.
The existence of this crossover illustrates how the velocity-dependent resetting dynamically suppresses large velocities as the force increases, effectively acting as a nonlinear friction. Importantly, this nonlinearity arises not from thermal fluctuations or interactions, but purely from the resetting mechanism. This provides a minimal and analytically tractable example of a system that exhibits an emergent nonlinear response even in the absence of noise. 
Since $F_c \propto \gamma^2/ \lambda$, stronger friction increases the crossover scale and extends the linear-response regime. Conversely, larger resetting rates lower the crossover threshold, making nonlinear effects prominent already at moderate driving strengths.
 
 \paragraph{Limiting Case 2: Diffusive dynamics with noise but no friction}

In the weak-field regime of Limiting Case 2, the stationary distribution $P_{\mathrm{st}}(v)$ can be expanded perturbatively in powers of the external force $F$. Retaining only the leading contribution, the mean velocity takes the linear-response form
\begin{equation}
  \langle v \rangle \sim \frac{3\,\mathrm{Ai}(0)}{2\,D^{1/3}\lambda^{2/3}}\, F, \qquad (F \to 0).
\end{equation}
Here, the value of the Airy function at the origin can be expressed as
\begin{equation}
  \mathrm{Ai}(0) = \frac{1}{3^{2/3}\Gamma(2/3)}.
\end{equation}
Thus, in contrast to the deterministic minimal model, the presence of diffusion restores linear response in the weak-driving limit. 
The proportionality coefficient defines an effective mobility, which depends nontrivially on the diffusion constant $D$ and the resetting rate $\lambda$, but remains finite. 
This indicates that thermal fluctuations regularize the singular nonlinear behavior observed in the absence of noise. 
We note that the coefficient obtained here does not exactly coincide with that predicted by Eq.~\eqref{eq: mean velocity general alpha=1}, which was derived using the simple shift approximation in the general weak-field analysis. 
The discrepancy reflects the fact that, in the frictionless case, the external driving slightly distorts the functional form of the stationary distribution rather than producing a pure shift.

In the strong-field regime, acceleration due to the external force dominates the dynamics, while fluctuations induced by diffusion become relatively negligible. In this limit, the system can be effectively regarded as deterministic by setting $D=0$. As a result, the mean velocity behaves as
\begin{equation}
  \langle v \rangle = \sqrt{\frac{2F}{\pi\lambda}}, \qquad (F \to \infty),
\end{equation}
recovering the square-root scaling found previously.

The crossover between linear and nonlinear response occurs around a characteristic force 
\begin{equation}
  F_c = \frac{8 D^{2/3}\lambda^{1/3}}{9\pi \left[\mathrm{Ai}(0)\right]^2}
\end{equation}
For \( F \ll F_c \), the system exhibits linear response, while for \( F \gg F_c \), the mean velocity grows sublinearly as \( \langle v \rangle \propto \sqrt{F} \). 
It is worth emphasizing that the dependence of the crossover force \( F_c \) on the resetting rate \( \lambda \) is qualitatively different between Limiting Case~1 and Case~2.
In Limiting Case1 (without diffusion), increasing $\lambda$ shifts the crossover to smaller forces, since stronger resetting enhances the nonlinear suppression of large velocities. In contrast, in Limiting Case2 (with diffusion), we find
\begin{equation}
F_c \propto \lambda^{1/3},
\end{equation}
so that larger resetting rates increase the crossover scale and extend the linear-response regime.
This opposite trend arises because diffusion regularizes the low-velocity behavior and modifies the weak-field mobility through the zero-field velocity fluctuations. While the asymptotic square-root scaling at large F remains universal, the location of the crossover is strongly affected by the interplay between diffusion and velocity-dependent resetting.

\begin{figure}
\includegraphics[width=.9\linewidth, angle=0]{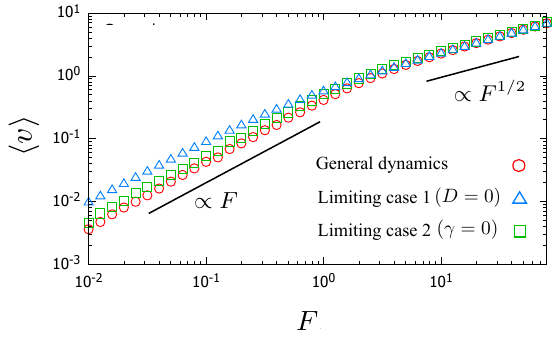}
\caption{
Mean velocity $\langle v \rangle$ as a function of the external force $F$ for the general dynamics ($D=1$, $\gamma=1$), together with the limiting cases $D = 0$ ($\gamma=1$) and $\gamma = 0$ ($D=1$).
Symbols represent numerical simulations, while the solid lines indicate the asymptotic scaling: linear response $\langle v\rangle \propto F$ for small $F$ and nonlinear scaling $\langle v\rangle \propto F^{1/2}$ for large $F$, given by Eqs.~\eqref{eq: mv small F case 2} and~\eqref{eq: mv large F case 2}.
}
\label{fig: mean velocity general}
\end{figure}

\subsection{Special Case: Constant-rate resetting (\( \alpha = 0 \))}

In this subsection, we analyze the steady-state mean velocity in the case \( \alpha = 0 \), where the resetting rate becomes independent of the particle's velocity.
This corresponds to a constant-rate (Poissonian) resetting process, widely studied in the literature.
 We focus on the mean velocity and evaluate it by considering the first moment of the steady-state velocity distribution.

The probability density $P(v,t)$ of the velocity $v$ obeys Eq.~(\ref{eq: master eq. general}) with $r(v)=\lambda$.
Multiplying the master equation by $v$ and integrating over the entire velocity space, we obtain an evolution equation for the mean velocity.
The drift term yields $F-\gamma\langle v\rangle$ after integration by parts, while the diffusion term vanishes under the assumed boundary conditions.
The loss term due to resetting contributes $-\lambda\langle v\rangle$, whereas the gain term proportional to $\delta(v)$ does not affect the mean velocity.
As a result, the time evolution of the mean velocity $\langle v\rangle$ is governed by 
\begin{equation}
\frac{d}{dt}\langle v\rangle = F-(\gamma+\lambda)\langle v\rangle .
\end{equation}
In the steady state, the left-hand side vanishes, leading to
\begin{equation}
\langle v\rangle_{\mathrm{st}} = \frac{F}{\gamma+\lambda}.
\end{equation}
This expression shows that, for $\alpha=0$, the equation governing the mean velocity is closed and does not involve higher-order moments of the velocity distribution.

From this result, it follows that the mean velocity responds linearly to the external force \( F \), with an effective friction coefficient given by \( \gamma + \lambda \), regardless of the strength of the field.
This behavior is in stark contrast to the case \( \alpha = 1 \), where the velocity-dependent resetting induces a nonlinear response and nontrivial asymptotic scaling.
Therefore, the constant-rate case \( \alpha = 0 \) serves as a useful reference for elucidating the effects of state-dependent resetting.

\section{Mechanism of Nonlinear Response}

To elucidate the origin of the nonlinear response observed in the preceding sections, we derive an exact moment balance relation that holds for the general dynamics, including friction and diffusion, with a velocity-dependent resetting rate $r(v) = \lambda |v|^{\alpha}$.
Multiplying the stationary master equation by $v$ and integrating over the entire velocity space, we obtain
\begin{equation}
\gamma \langle v \rangle
=F-
\lambda \langle v |v|^{\alpha} \rangle .
\label{eq: moment balance general}
\end{equation}
This exact relation expresses a balance between external driving $F$, viscous dissipation $\gamma \langle v\rangle$, and resetting-induced dissipation $\lambda \langle v |v|^{\alpha} \rangle$. 
Importantly, Eq.~(\ref{eq: moment balance general}) does not depend on the specific values of $D$, $\gamma$, or $\alpha$, and therefore provides a general framework for understanding the emergence of nonlinear response in resetting-driven transport processes.

\subsection{Weak-field regime}

In the weak-driving limit, the stationary distribution remains close to its zero-field form and is only weakly asymmetric. 
The scaling of the mixed moment depends on the sign of the exponent $\alpha$.

For $\alpha>0$, the factor $|v|^{\alpha}$ is regular near $v=0$, so that the mixed moment is controlled by the typical velocity scale of the zero-field stationary distribution rather than by a singular contribution from the origin. 
In the weak-field regime, the stationary distribution remains close to its zero-field form and is only weakly shifted by the external force. 
Writing this shift as
\begin{equation}
P_{\mathrm{st}}(v)\simeq P_0(v-\delta)\simeq P_0(v)-\delta\,\frac{dP_0(v)}{dv},
\end{equation}
with $\delta \sim \langle v\rangle$, and expanding
\begin{equation}
(v+\delta)|v+\delta|^{\alpha}
=
v|v|^{\alpha}
+
(\alpha+1)\delta |v|^{\alpha}
+
O(\delta^2),
\end{equation}
one finds that the leading odd contribution is linear in the shift. 
As a result, the mixed moment scales linearly with the mean velocity,
\begin{equation}
\langle v |v|^{\alpha} \rangle
\approx
C_{\alpha}\langle v\rangle ,
\end{equation}
where
\begin{equation}
C_{\alpha}=(\alpha+1)\langle |v|^{\alpha}\rangle_{0}
\end{equation}
is determined by the zero-field stationary distribution.
Substituting this relation into Eq.~\eqref{eq: moment balance general} yields
\begin{equation}
\langle v\rangle
\simeq
\frac{F}{\gamma+\lambda C_{\alpha}},
\label{eq: linear response general}
\end{equation}
showing that resetting contributes an effective linear dissipation and the system exhibits conventional linear response for sufficiently small $F$. 
Importantly, the linear-response regime discussed above exists only when a regular dissipative scale (such as $\gamma$ or diffusion-induced width) provides a finite velocity scale in the stationary state.

For $\alpha<0$, however, the factor $|v|^{\alpha}$ enhances the contribution of small velocities. 
In this case the mixed moment is dominated by the behavior of the distribution near $v=0$, and the simple linear closure used above no longer applies. 
Instead, in the weak-field regime the stationary distribution remains narrow and is characterized by a single velocity scale $v_*$ set by the mean velocity, $v_* \sim \langle v\rangle$. 
The mixed moment then scales as
\begin{equation}
\langle v |v|^{\alpha} \rangle \approx v_*^{\alpha+1} \approx \langle v\rangle^{\alpha+1},
\end{equation}
since $v|v|^{\alpha}$ has scaling dimension $\alpha+1$ with respect to the characteristic velocity. 
Substituting this scaling into Eq.~\eqref{eq: moment balance general} gives
\begin{equation}
\langle v \rangle \propto F^{1/(\alpha+1)},
\end{equation}
indicating a superlinear response in the weak-field regime for $\alpha<0$.

\subsection{ Strong-field regime}

In the strong-field regime the velocity distribution becomes strongly skewed and is characterized by a single velocity scale set by the mean velocity $\langle v\rangle$. 
Consequently,
\begin{equation}
\langle v|v|^{\alpha}\rangle
\sim
\langle v\rangle^{\alpha+1}.
\end{equation}
This scaling reflects the fact that resetting events predominantly occur at velocities of order $\langle v\rangle$, so that the resetting-induced dissipation grows as a power of the characteristic velocity of the system. 
As a result, the resetting mechanism acts as an effective nonlinear friction that increases with velocity when $\alpha>0$ and decreases with velocity when $\alpha<0$.

For $\alpha>0$, the resetting rate increases with velocity and therefore remains relevant even at large driving. 
Equation~\eqref{eq: moment balance general}  then reduces asymptotically to
\begin{equation}
F
\sim
\lambda \langle v \rangle^{\alpha+1}\quad (F\to\infty),
\end{equation}
which yields
\begin{equation}
\langle v \rangle
\propto
F^{1/(\alpha+1)} \quad (F\to\infty).
\label{eq: strong limit general}
\end{equation}
Nonlinear response is generally associated with the emergence of higher-order moments or nonlinear susceptibilities beyond the linear regime.
In the present model, this structure appears explicitly in the exact moment balance relation, where the mean velocity couples directly to the mixed moment $\langle v |v|^{\alpha} \rangle$.
As a result, velocity-dependent resetting acts as a nonlinear dissipation mechanism at strong driving, producing sublinear response for $\alpha>0$ and superlinear response for $\alpha<0$.
Importantly, the exponent $1/(\alpha+1)$ is independent of friction and diffusion, reflecting the dominance of the resetting mechanism in setting the transport scale in this regime.

For $\alpha<0$, the resetting rate decreases with velocity, so that
$r(v)\to0$ as $v$ becomes large.
In the strong-field regime the mixed moment
$\langle v|v|^{\alpha}\rangle$
therefore becomes negligible compared with $F$,
and Eq.~\eqref{eq: moment balance general} reduces to
\begin{equation}
\gamma\langle v\rangle \simeq F .
\end{equation}
Hence the response asymptotically approaches the linear form
$\langle v\rangle \sim F/\gamma$ at sufficiently large driving.

\subsection{ Deterministic limit as a singular case}

In the absence of friction and diffusion ($\gamma = D = 0$), Eq.~\eqref{eq: moment balance general} reduces exactly to
\begin{equation}
F
=
\lambda \langle v |v|^{\alpha} \rangle .
\end{equation}
Since no linear dissipative term is present, the balance between driving and resetting holds at all force strengths. Consequently,
\begin{equation}
\langle v \rangle
\propto
F^{1/(\alpha+1)}
\end{equation}
for all $F$, and no linear regime exists even in the weak-driving limit.
The deterministic model therefore represents a singular limit in which nonlinear response appears at leading order, rather than emerging as a higher-order correction.

\subsection{Physical Interpretation}

The essential mechanism can be summarized as follows:
\begin{enumerate}
\item[(i)] Near equilibrium, resetting contributes effectively as a linear friction.
\item[(ii)] Far from equilibrium, resetting suppresses large velocities through a nonlinear feedback proportional to $|v|^{\alpha}$.
\item[(iii)] When ordinary dissipation is absent, this nonlinear feedback dominates at all driving strengths.
\end{enumerate}
Importantly, the breakdown of linear response does not originate from non-Markovian memory, quenched disorder, or many-body interactions. 
The underlying dynamics remains strictly Markovian. 
The nonlinearity emerges directly from the state-dependent resetting rate, which couples the transport coefficient to higher-order velocity moments through the exact moment balance relation.
Thus, linear response fails not because of dynamical complexity, but because the effective dissipation itself becomes nonlinear in the instantaneous state of the system.

\section{Conclusion and Outlook}

In this work, we have demonstrated that state-dependent stochastic resetting can fundamentally alter the response properties of driven steady states. 
By analyzing a minimal yet analytically tractable model, we established an exact moment balance relation that unifies the roles of external driving, viscous dissipation, and resetting-induced dissipation. 
This relation reveals that the breakdown of linear response is not an incidental feature of a particular limit, but a structural consequence of velocity-dependent resetting.

Our analysis shows that when the resetting rate depends on velocity as $r(v)=\lambda |v|^{\alpha}$, the mixed moment $\langle v |v|^{\alpha} \rangle$ controls the nature of the steady-state response. 
In the weak-driving regime and in the presence of regular dissipation, resetting contributes effectively as an additional linear friction, leading to conventional linear response. 
However, at strong driving, or in the absence of ordinary dissipation, the resetting term scales nonlinearly with the typical velocity, yielding the exact power-law behavior $\langle v \rangle \propto F^{1/(\alpha+1)}$. 
In the deterministic limit, this nonlinear scaling persists for all $F$, demonstrating that the linear-response regime itself can disappear as a singular limit of the dynamics.

These results identify state-dependent resetting as a minimal mechanism capable of generating intrinsic nonlinear response without interactions, disorder, or memory effects. 
In contrast to many previously studied systems where nonlinear transport arises from collective effects, spatial heterogeneity, or anomalous statistics, the present mechanism emerges purely from a state-dependent dynamical constraint. 
The appearance of square-root scaling for $\alpha=1$ illustrates how a simple velocity-dependent resetting rule can produce nonanalytic transport behavior even in a fully solvable model.

More broadly, our findings suggest that response properties in nonequilibrium systems can be controlled not only by thermodynamic forces and fluctuations, but also by dynamical rules that couple transition rates to the system’s instantaneous state. 
This perspective naturally connects to stochastic thermodynamics, where nonequilibrium dynamics is formulated in terms of transition rates that break detailed balance and where entropy production and transport can be analyzed systematically~\cite{baiesi2009nonequilibrium,Baiesi2009,Baiesi2013,seifert2012stochastic,Hatano2001}.
Thermodynamic aspects of resetting processes have also been explored~\cite{fuchs2016stochastic,Mori2023}.

Future directions include extending the present framework to higher dimensions, interacting particles, or resetting protocols that depend on energy rather than velocity. It would also be of interest to quantify entropy production and fluctuation relations in velocity-dependent resetting systems, and to explore possible experimental realizations in cold atoms, granular media, or active-matter systems with state-dependent scattering mechanisms.

We expect that the mechanism uncovered here will stimulate further studies on how dynamical constraints govern the emergence, recovery, and universality of linear response far from equilibrium.

\section*{Acknowledgement}
T.A. was supported by JSPS Grant-in-Aid for Scientific Research (No.~C 21K033920).



\bibliography{akimoto}

\end{document}